\documentclass{kapproc} 
\usepackage[dvips]{graphicx}
\upperandlowercase
\kluwerbib 

\begin{document}

\articletitle[Intermediate Polars in Low States]
{Intermediate Polars in Low States}

\author{Brian Warner}

\altaffiltext{}{Department of Astronomy,
University of Cape Town, Rondebosch 7700, South Africa}
\email{warner@physci.uct.ac.za}

\begin{abstract}
Although no intermediate polar (IP) has been observed in recent 
years to descend into a state of low rate of mass transfer, there are 
candidate stars that appear to be already in intermediate and low 
states. V709 Cas is probably an intermediate state IP; NSV 2872 
appears to be a low state IP, and will probably resemble the long 
orbital period IP V1072 Tau if it returns to a high state. The 
enigmatic star V407 Vul, which has had many interpretations as an 
ultra-short period binary, has many resemblances to the pre-cataclysmic 
variable V471 Tau, and may therefore be an IP 
precursor of quite long orbital period, and not yet a fully fledged 
cataclysmic variable.
\end{abstract}


\section{Introduction}

The two principal classes of magnetic cataclysmic variable (mCV) 
are the synchronous rotators and the asynchronous rotators. The 
former, known as polars, possess white dwarf primaries with fields 
strong enough to interact with the secondary star and effect 
synchronous rotation of the primary with the orbital motion of the 
secondary. In the latter, known as intermediate polars (IPs) the 
primary is less strongly magnetized so the primary rotates with a 
period different from $P_{orb}$.

    The polars are well known for having unstable mass transfer 
from the secondary to the primary -- with resulting high and low 
states of accretion luminosity. Although relatively low states of IPs 
are known from studies of archived plates (Garnavich \& Szkody 
1988), and V1223 Sgr in particular has had extensive  low states, it 
is a curious, and often regretted, fact that since they were 
recognized in the early 1980s, none of the well-studied IPs has 
entered into a low state. There would be obvious advantages to 
studying IPs in low states, particularly if the rate of mass transfer 
$\dot{M}$ were to shut off completely -- the spectrum of the white 
dwarf would then be visible uncontaminated by accretion emission 
and the possibility of measuring field strengths via the Zeeman 
effect, as is done during polar low states, would appear.

\section{V709 Cas}

    A first move in this direction has been made possible by V709 
Cas, which has $P_{orb}$ =  5.34 h and $P_{rot}$ = 5.22 min. The optical 
spectrum of this system shows the presence of broad absorption 
lines of the white dwarf primary, on which are superimposed the 
continuum and emission lines from the accretion disc and accretion 
curtains (Bonnet-Bidaud et al.~2001). Only an upper limit $B < 10$ 
MG is so far possible for the field strength, but variable X-Ray 
flux on time scales of months to years show that V709 Cas has 
unstable $\dot{M}$ which may at some time fall low enough for the 
white dwarf spectrum to be more clearly studied. Both Bonnet-
Bidaud (2001) and De Martino et al.~(2001), from different 
approaches, find $\dot{M} \sim 1 \times 10^{16}$ g s$^{-1}$, 
which is a factor of 4 -- 10 lower than the values for $\dot{M}$ 
deduced for more normal IPs (see Table 7.4 of Warner 1995).

   Thus V709 Cas appears to be an IP in an intermediate state of 
Mdot, which is compatible with the weakly visible spectrum of the 
primary. Its present brightness is $m \sim 14$, so a reconstruction of its 
historic light curve from archived plates is both feasible and 
desirable -- if it attains a normal high state it could reach 12th 
magnitude.

\section{V1072 Tau and NSV 2872}

As a second step we might wonder whether there are in fact IPs in 
long-lived low states, already known but not properly recognized 
as such. As an example we can consider the IP V1072 Tau and ask 
what it would look like if it were to descend to a state of low $\dot{M}$.

   V1072 Tau is a long period IP, with $P_{orb}$ = 9.95 h and $P_{rot}$ = 62.0 
min (Remillard et al.~1994). Its optical spectrum is that of a late K 
star with superimposed emission lines and continuum from the 
accretion process. The K spectral type of the secondary is 
appropriate for a Roche lobe filling star in long period orbit. If the 
system were in a low state the spectrum would be dominated by 
the K star absorption spectrum, but there would be a contribution 
at short wavelengths (the U band, and shorter wavelengths), with 
the flux modulated at 62 min period either from the low level of 
Roche lobe overflow, or from magnetically channeled wind from 
the secondary (as in V471 Tau, which we discuss below).

   Such an object might be quite difficult to discover (but would 
certainly be found eventually in any wide field survey like the 
Sloan Digital Sky Survey). It could, however, be readily found if 
the long-term light curve shows higher states of $\dot{M}$.

   The object described here sounds very like the recently 
recognized low state CV known as NSV 2872. This star was first 
found to be variable by Ruegemer (1933), and a more complete 
light curve was provided by Zinner (1932) and Florja \& Kukarkin 
(1935). It has a brightness range of $m_{pg}$ = 11.4 -- 14.5 and was 
classified in the GCVS as a suspected nova-like or dwarf nova. A 
spectrum obtained near the lower end of its brightness range 
showed only the absorption lines of an early K star, with no 
emission lines, and was consequently thought not to be a CV (Liu 
\& Hu 2000). However, Kozhevnikov (2003) has made extensive 
photometric observations near minimum brightness (at $B \sim 14.4$: 
Kozhevnikov, private communication) and concludes that it (a) has 
very low level rapid flickering, characteristic of a CV in a low state 
of $\dot{M}$ and (b) has a coherent periodicity at 87.850 min of low 
and variable amplitude (3 -- 8 mmag).

     NSV 2872 therefore looks very much what we would expect of 
an IP in a long orbital period, accreting at a very low rate. The 
already recognized range of brightness of $\sim$ 3 mag shows that it 
can have high states. This is another system for which a more 
complete historical light curve would be very useful -- and one that 
should be monitored regularly in order to detect the occurrence of 
a high state, where it could take on the appearance of V1072 Tau. 
A far UV study of the hot component is also obviously of some 
interest.

\section{V471 Tau}

   The eclipsing binary V471 Tau is a member of the Hyades 
cluster and is a detached pair consisting of a 35\,000 K white dwarf 
in orbit around a dwarf K2 secondary with $P_{orb}$ = 12.5 h. What 
makes it more interesting is that it shows a 9.25 min modulation in 
the U band with an amplitude of 9.5 mmag (Clemens et al.~1992) 
that is $180^{\circ}$ out of phase with pulsed soft X-Rays at the same 
period. From the known contribution of the white dwarf to the U 
band we can estimate that the true amplitude of the 9.25 min 
modulation is $\sim$ 30 mmag.

   The explanation of the modulation is that the white dwarf has a 
magnetic field sufficiently strong to capture and channel part of the 
stellar wind from the secondary (Clemens et al.~1992). Thus V471 
Tau is really a pre-IP, which will become an IP as soon as the 
secondary begins to overflow its Roche lobe.

   The optical spectrum of V471 Tau shows the continuum and 
absorption lines of a K2 dwarf, currently with no emission lines 
(though in past years there has been H$\alpha$ in emission, around phase 
0.5 of the orbit) -- Rottler et al.~(1998). In the UV, of course, a 
continuum contribution from the white dwarf is present.

   There are strong similarities between V471 Tau and NSV 2872 -- the 
difference being that the former does not have the low level 
flickering that is probably the signature of Roche lobe overflow.

\section{V407 Vul}

  V407 Vul has had a roller coaster career. Discovered initially as 
an X-Ray source, RX\,J1919.4+2456, modulated at 569 s period 
and thought to belong to the class of soft X-Ray IPs (Haberl \& 
Motch 1995; Motch et al.~1996), it was later suggested that it could 
be a strongly magnetic He-transferring system (i.e., an AM CVn 
polar) where the observed period would be orbital (Cropper et al.~1998). 
The optical component was later identified, at $I \sim 18.6$, and 
was found to be modulated at the 9.5 min period, being in 
antiphase with the X-Ray variation (Ramsay et al.~2000). The X-Ray 
flux was found to vary by an order of magnitude on a time 
scale of a year. Later observations (Ramsay et al.~2002) showed 
modulation in the $R$ and $V$ bands, the latter with an amplitude of 
$\sim 65$ mmag, and a spectral flux distribution like that of a reddened K 
star, with no emission lines.

    Alternate models were proposed for V407 Vul, to account for 
the lack of emission lines and absence of polarization. Marsh \& 
Steeghs (2002) suggested that the inter-star stream of gas impacts 
directly onto the primary, as in an Algol system, thus preventing an 
accretion disc from being formed. A unipolar-inductor model, as in 
the Jupiter-Io system, was proposed by Wu et al.~(2002).

   With the acquisition of a higher quality spectrum of V407 Vul it 
is now known that the spectrum is simply that of an apparently 
normal reddened K star, with the addition of a featureless pulsed 
component at the shortest wavelengths (Steeghs 2003). There are 
no emission lines and there is no helium present.

    The similarities of the spectrum, X-Ray and pulsation period of 
V407 Vul and V471 Tau are quite striking. There are parallels, too, 
with the properties of NSV 2872.

\section{Conclusions}

   Although no recognized IP has entered a low state since this 
class of mCV was identified, there appear to be low $\dot{M}$ IPs 
available for study. V709 Cas has the signature of an IP in an 
intermediate state, and will repay study if it moves up or down 
from that state. NSV 2872 has features that would be expected of a 
long orbital period IP in a low state -- in particular, if it returns to a 
high state it should resemble the high $\dot{M}$ IP V1072 Tau.

   The enigmatic object V407 Vul, which has had many 
interpretations, has a strong resemblance to V471 Tau and so 
strictly is probably not a CV at all -- it may be a long period pre-CV 
which will become an IP. Such systems may be relatively 
common -- they are not easy to detect optically because they are 
dominated by the K star flux, yet the brightest is as close as the 
Hyades cluster. Lower mass systems, with M spectral type 
companions, will be easier to find as White Dwarf/M dwarf pairs, 
but will require appropriate extended photometry to detect 
rotationally modulated flux. Similarly, new detections through 
discovery of periodically modulated X-Ray flux will require 
extended pointed observations.

\begin{acknowledgments}
I thank Dayal Wickramasinghe for helpful discussions. My 
research is funded by the University of Cape Town.
\end{acknowledgments}


\begin{chapthebibliography}{1}
\bibitem{bb01}   Bonnet-Bidaud, J.M., Mouchet, M., de Martino, D., Matt, G. \& 
                 Motch, C. (2001). A\&{A}, 374, 1003
\bibitem{cl92}   Clemens, J.C., et al. (1992). ApJ, 391, 773
\bibitem{cr98}   Cropper, M., et al. (1998). MNRAS, 293, 57L
\bibitem{dm01}   de Martino, D., et al. (2001). A\&{A}, 377, 499
\bibitem{f35}    Florja, N.F. \& Kukarkin, B.V. (1935). Perem. Zvezdy, 5, 19
\bibitem{gs88}   Garnavich, P. \& Szkody, P. (1988). PASP, 100, 1522
\bibitem{hm95}   Haberl, F. \& Motch, C. (1995). A\&{A}, 397, L37 
\bibitem{ko03}   Kozhevnikov, V.P. (2003). A\&{A}, 398, 267 
\bibitem{li00}   Liu, W. \& Hu , J.Y. (2000). ApJS, 128, 387
\bibitem{ms02}   Marsh, T.R. \& Steeghs, D. (2002). MNRAS, 331, 7L
\bibitem{mo96}   Motch, C., et al. (1996). A\&{A}, 307, 459
\bibitem{ra00}   Ramsay, G., Cropper, M., Wu, K., Mason. K.O. \& Hakala, P. 
                 (2000). MNRAS, 311, 75
\bibitem{ra02}   Ramsay, G., et al. (2002). MNRAS, 333, 575
\bibitem{re94}   Remillard, R.A., et al. (1994). ApJ, 428, 785
\bibitem{ro98}   Rottler, L., Batalha, C., Young, A. \& Vogt, S. (1998). 
                 BAAS, 192, 6720
\bibitem{ru33}   Ruegemer, H. (1933). AN, 248, 410
\bibitem{st03}   Steeghs, D. (2003). Workshop on Ultracompact Binaries, Santa 
                 Barbara. \\
                See {\tt http://online.kitp.ucsb.edu/online/ultra$\_$c03/steeghs/}
\bibitem{wa95}   Warner, B. (1995). Cataclysmic Variable Stars. 
                 Cambridge University Press 
\bibitem{wu02}   Wu, K., Cropper, M., Ramsay, G. \& Sekiguchi, K. (2002). 
                 MNRAS, 331, 221
\bibitem{zi32}   Zinner, E. (1932). AN, 246, 17
\end{chapthebibliography}
\end{document}